\documentclass[prl,aps,twocolumn,showpacs]{revtex4}
\usepackage{amsmath,graphicx,epsfig}
\begin{document}

\def\pd#1#2{\frac{\partial #1}{\partial #2}}

\title{\bf Vortex nucleation by collapsing bubbles in Bose-Einstein condensates}
\author {Natalia G. Berloff${}^1$ and Carlo F. Barenghi${}^2$}
\affiliation {${}^1$Department of Applied Mathematics and Theoretical Physics,
University of Cambridge, Wilberforce Road, Cambridge, CB3 0WA\\
${}^2$School of Mathematics and Statistics, University of Newcastle,
  Newcastle upon Tyne NE1 7RU\\ 
}
\date {January 3, 2004}

\begin {abstract} The nucleation of vortex rings accompanies the collapse
  of ultrasound bubbles in superfluids. Using the Gross-Pitaevskii
  equation for a uniform condensate we elucidate the various
  stages of the collapse of a stationary spherically symmetric
  bubble and establish conditions necessary for  vortex
  nucleation. The
  minimum radius of the stationary bubble, whose collapse leads to vortex
  nucleation, was found to be $28\pm 1$ healing lengths. The time after
  which the nucleation becomes possible is determined as a function of
  bubble's radius. We show that  vortex nucleation takes place in
  moving bubbles of even smaller radius if the motion made them
  sufficiently oblate.

  \end{abstract}
\maketitle

In this Letter we establish a new mechanism of vortex nucleation
in a uniform condensate. Previously, the nucleation of vortices in a uniform
condensate  has been connected to
 critical velocities \cite{frisch,br7,br8}, instabilities of the
initial states \cite{berloff} or to a transfer of energy among the
solitary waves \cite{pade}. Moving positive \cite{br7} and negative
\cite{br8} ions  were shown to generate vortex rings on their surface
where the speed of sound was exceeded. 
Experiments in superfluid helium have demonstrated long time ago the production
of quantised vortices and turbulence \cite{careySchwarz} by the collapse of 
cavitated bubbles \cite{finch} generated by ultrasound in the megahertz 
frequency range. The aim of this Letter is to analyse theoretically
for the first time the physics of this process in the context of the
Gross-Pitaevskii (GP) equation. Vortex nucleation by collapsing 
bubbles could also be studied in the context of (non-uniform) 
atomic condensates (BEC), for which the GP equation provides a
quantitative model,
thus providing experimentalists with a new mechanism to produce
vortices in BEC systems, alongside rotation \cite{madison},
the decay of solitons \cite{anderson} and phase imprinting \cite{leonhardt}.
Moreover, our work illustrates a new aspect of vortex-sound interaction
in a Bose- Einstein condensate,
a topic which is receiving increasing attention \cite{kivotidesEtAl}.

We write the GP equation in dimensionless form as
\begin{equation}
-2{\rm i} \pd \psi t =  \nabla^2 \psi +(1- |\psi|^2-V({\bf x},t)) \psi,
\label{gp}
\end{equation}
in dimensionless variables
such that the unit of length corresponds to the healing
length $\xi$, the speed of sound is $c=1/\sqrt{2}$,  and the density at
infinity is $\rho_\infty=|\psi_\infty|^2=1$. To convert the
dimensionless units into values applicable to superfluid helium-4, we
take the number density as $\rho=2.18\times 10^{28} {\rm m}^{-3}$, the
quantum of circulation as $\kappa=h/m=9.92\times 10^{-8} {\rm m^2
  s^{-1}}$, and the healing length as $\xi=0.128 {\rm nm}$. This gives
  a time unit $2\pi\xi^2/\kappa\sim1 {\rm ps}$. Whereas for a sodium
  condensate with $\xi\approx 0.14\mu {\rm m}$, the time unit is about
  $8 {\rm ns}$.
$V({\bf x},t)$
is the
potential of interaction between a boson and a bubble. We will assume
 that the bubble acts as an infinite potential barrier to the
 condensate, so that no bosons can be found inside the bubble
 ($\psi=0$) before the collapse. This is achieved by setting $V$ to be
 large inside the bubble and zero outside.

First we consider the case of a stationary spherically symmetrical
bubble. The spherical symmetry allows us to reduce the problem to
dimension one,
so that  the equation (\ref{gp}) for 
$\psi=\psi(r,t)$
becomes 
\begin{equation}
-2 {\rm i} \psi_t = \psi'' + 2 \psi'/r + (1 - |\psi|^2) \psi,
\label{gp1d}
\end{equation}
where $r^2=x^2+y^2+z^2$. Equation (\ref{gp1d}) is numerically
integrated using fourth order finite differences discretization
in space and fourth order Runge-Kutta method in time.
Before the collapse the field around the bubble of radius $a$ is
stationary, $\psi_t=0$.  The boundary conditions are 
$\psi(a,t)=0$
stating that the bubble surface is an infinite potential barrier to
the condensate and 
$\psi(\infty,t)=1$.
The stationary solutions for
various $a$  were found by
the Newton-Raphson iterations. The solutions are
$\psi(r)=(0,0)$ if $r\le a$ and $\psi(r)=(R_a(r),0)$ if $r > a$,
with the graphs of $R_a(r+a)$ for $a=1,2,10,30$ given on Figure 1. If the radius of the bubble, $a$, is
sufficiently large, then we can set $r=a+\xi$ and to the leading order
get $R''(\xi)+[1 -R(\xi)^2]R(\xi) =0$ which has the  solution, satisfying the
boundary conditions,  $R(\xi)=\tanh(\xi/\sqrt{2})$. The total energy of the
system \cite{jr}, ${\cal E} = \tfrac{1}{2}\int |\nabla \psi|^2 \,
dV + \tfrac{1}{4}\int(1-|\psi|^2)^2\, dV$ depends on the radius of the
bubble, and therefore,  on the form of $R_a$:
\begin{equation}
{\cal E}=\frac{\pi a^3}{3}+2\pi\int_a^\infty [R_a'(r)^2 +
\tfrac{1}{2}(1 - R_a(r)^2)^2]r^2\, dr.
\label{e}
\end{equation}
The insert on Figure 1 shows the loglog plot of the energy vs radius
of the bubble together with the linear fit. For $a>20$ the energy depends
on the radius as ${\cal E} \sim 1.65\,a^{2.9}$.
\begin{figure}
\caption{(colour online) The plot of the amplitude of the  solution around the
  stationary bubbles of radii $a=1$ (red),$2$ (green), $10$ (blue) and
  $30$ (black) (the smaller $a$
  corresponds to a  steeper amplitude). The loglog plot of the energies
  of the solutions with various $a$ are shown on the inset together
  with the linear fit. 
}
\centering
\bigskip
\bigskip
\epsfig{figure=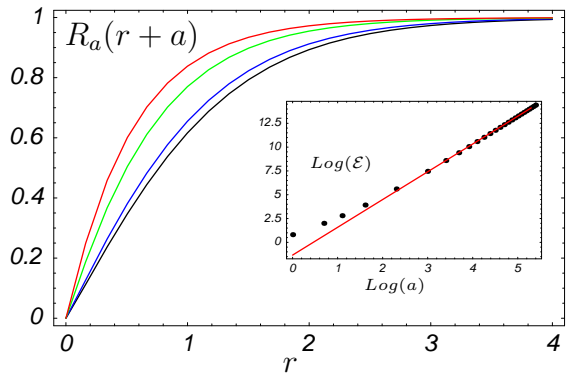, height = 2 in}
\end{figure}
From the energy conservation it is clear that after the bubble
collapses and the condensate fills the cavity the necessary (but not
sufficient) condition for vortex nucleation is that the energy has to
be greater than that of one vortex ring. The minimal energy of the
vortex solution was found in \cite{jr} to be about ${\cal E} \sim
  55 \pm 1$ which corresponds to the minimum radius of $a=2.2$ with
  ${\cal E}=55.7$. As the condensate fills  the cavity,
  most of the energy will be emitted via the sound waves, so the
  energy of the bubble has to be sufficiently greater than the
 energy of a single vortex ring to allow for such an emission.

The time-dependent evolution of the condensate after the bubble
collapses  involves several stages. The overall picture is complicated
by a complex interplay between  dispersive and nonlinear
effects. Dispersive effects become important on the wavelengths of
order of the healing length with the group velocity approximately
given by $\partial (\sqrt{k^2/2+k^4/4})/\partial k$ for the
perturbation propagating along the uniform state $\psi=1$ towards
infinity and with the
group velocity approximated by
$\partial (k^2-1)/2\partial k$ for the perturbation moving along the
uniform state $\psi=0$ towards the centre of the cavity.  The wavetrain
generated by the nonlinearity is moving slower with the larger
wavelengths than the dispersive wavetrain. During the first stage 
dispersive and nonlinear wavetrains are generated near the areas of the largest
curvature of the stationary profile.
The Fourier components propagate at different velocities generating
 wave packets moving in opposite directions. This stage of the
 evolution is characterised by a flux of particles towards the centre
 of the cavity as the oscillations of the growing amplitude are being
 formed on the real  and imaginary parts of the wave function and the
 slope of the steep density front  is getting smaller; see
 Figure 2. 
%
%
%
%
\begin{figure}
\caption{(colour online) The plot of the density of the condensate as
  a function of the distance from the centre of
  the cavity
   for  $t = 0$ (black), $10$ (blue), $20$ (green), $30$ (red), $40$ (magenta) after the collapse of the bubble of
  the radius $a=50$. }
\centering
\bigskip
\bigskip
\epsfig{figure=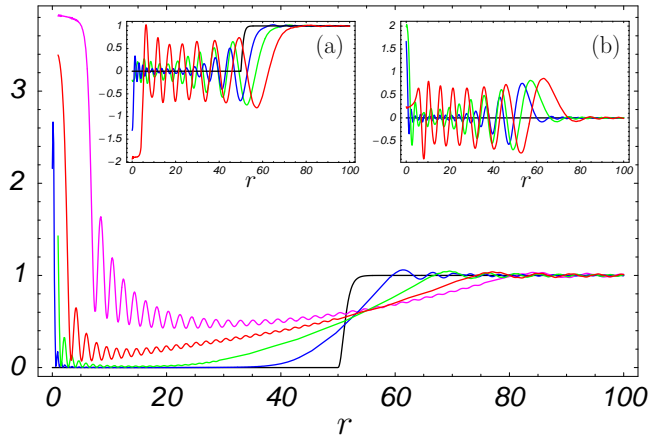, height = 2.3 in}
\end{figure}
To follow the evolution of the particle flux we calculated the  density
per unit volume averaged over spheres of different radii, centred at the origin,
as functions of time. Figure 3 shows these functions for the radius of
the cavity $a=128$ and the radii of the spheres  over which the
averaging of the density is performed being $4, 8, 16, 32,$ and
$64$. From Figure 3 it is clear that there is a definite moment of
time $t^*\approx 97$ when the density per unit volume reaches its maximum at
the same time (and taking the same value) for two smallest spheres. This moment indicates the start
of a qualitatively new stage of the evolution in which there is an
outward flux of particles as the condensate that overfilled the cavity
began to expand.  The density gradually
approaches the uniform state $\rho=1$.
\begin{figure}
\caption{(colour online) The plots of the density per unit volume as function of time
  for various radii of spheres over which the averaging is
  performed. The initial radius of the cavity 
  in this case
  is $a=128$. The radii of
  averaging spheres are $b=4,8,16,32, 64$. The average density is
  calculated as $\bar{\rho}_{b} = 3\int_0^b\rho(r)r^2\,dr/b^3.$
}
\centering
\bigskip
\bigskip
\epsfig{figure=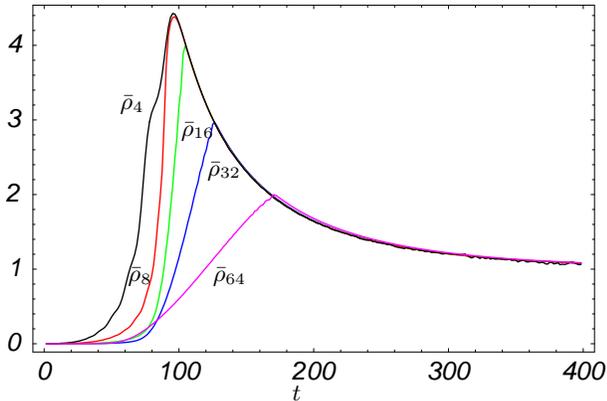, height = 2 in}
\begin{picture}(0,0)
\put(-125,-5) {$t$}
\put(-190,105) {$\bar{\rho}_{4}$}
\put(-187,40) {$\bar{\rho}_{8}$}
\put(-167,95) {$\bar{\rho}_{16}$}
\put(-157,80) {$\bar{\rho}_{32}$}
\put(-155,40) {$\bar{\rho}_{64}$}
\end{picture}
\end{figure}

 It is after the time $t^*$ that we expect the instability to set
in and to give rise to vortex nucleation as the outward flow can
support outward moving vortex rings. We calculated $t^*$ for various
radii of the bubble  and determined that for $a>10$ the time of the
start of the outflow from the centre of the cavity is approximated
quite well by the linear function $t^*\sim 1.96 + 0.72\,a.$

As it seen on the insets of Figure 2, some of the  surfaces of zero real
and zero imaginary parts may be just a healing length apart
(for instance, at $t=100$ for the radius of the cavity $a=128$, one of the  surfaces of zero of the real part has radius
$16.02$ and the nearest surface of the zero of imaginary part has
radius $17.11$). Small radial perturbations on these zero surfaces can lead to their
overlap  which creates topological zero curves of the wavefunction which 
quickly adjust their form by emitting sound waves to become axisymmetrical
vortex rings.

The smaller $a$ is, the larger the distances between the zero
surfaces are during the condensate expansion, so it would require a much
larger perturbation to bring these surfaces to intersect. In this case
the instability mechanism would be somewhat different taking longer
time to develop. The radial 'dips' of the density of the expanding
condensate, noticeable in Figure 2, are unstable to non-spherically
symmetrical perturbations, similar to the instability of the
Kadomtsev-Petviashvili 2D solitons in 3D \cite{berloff}. Depending on the energy
carried by these 'dips,' they evolve into either vortex solutions or 
 sound waves. From these considerations we expect three possible
 outcomes after bubble collapses: (1) if the radius of the bubble is
 smaller then some critical radius $a^*$, the density 'dips' generated
 by the expanding condensate have rather small amplitude
 that decreases even further as they travel away from the centre
 quickly becoming sound waves before the instability has time to
 develop; (2) after the collapse of a bubble of an intermediate size,
 say, of the radius  $a^*< a < \hat a$, the waves of  sufficiently large amplitudes are
 generated and the instability of these waves develops in time
 inversely proportional to the radius $a$; (3) if the radius is
 sufficiently large, $a > \hat a$, the time of the first  vortex nucleation is
 approximately 
 given by the moment of the start of the outward flux of the particles
 $t^*$ approximated above. As a condensate continues to
 expand the instability mechanism described in (2) is further facilitated by
 the broken symmetry resulting from the previous nucleation events
 which leads to even more vortex rings being nucleated.

To confirm the scenario outlined above we performed full three-dimensional 
calculations 
for cavities of various radii in a 
computational box of the side $200$ healing lengths \cite{numer2}. We determined
that there is a critical radius of the bubble for which vortex ring
nucleate $a^*\sim 28 \pm 1$. The borderline radius between regime
(2) and (3) was found as $\hat a\sim 45$.  Figure 4 shows the density isoplots
at the various time snapshots after the bubble of radius $a=50$
collapsed. Notice, that the non-symmetry of the field  that was
  created at the time of nucleation ($t\approx 40$) continues to produce even more vortex rings as
  condensate expands. Each vortex ring at the moment of its birth has
  zero radius, as the surfaces of zero real and imaginary parts touch
  each other, and gradually evolve into a vortex ring of increasingly larger and radius, as clearly seen on Figure 4. The process in which
  solitary waves evolve into states of a higher energy was elucidated
  in \cite{pade}. A finite amplitude sound wave that moves behind
   a vortex ring transfers its energy to it, allowing the vortex ring
   to grow in size. The radius of the vortex ring stabilises only when
   it  travelled sufficiently far from the center of the collapsing
   bubble, where the flow became almost uniform. The larger the radius
   of the bubble, the more finite amplitude sound waves will be generated
   at shorter distances, the larger the size of the final ring is
   going to be.

\begin{figure}
\caption{(colour online) Time snapshots of the density isoplots $\rho=0.1$ after the
  collapse of the bubble of radius $a=50$. The side of the
  computational box is $200$ healing lengths and the distance between
  ticks on the side corresponds to $20$ healing lengths. The vortex
  rings nucleate at about $t\sim 40$ (compared with $t^*\sim 38$).
}
\centering
\bigskip
\bigskip
\epsfig{figure=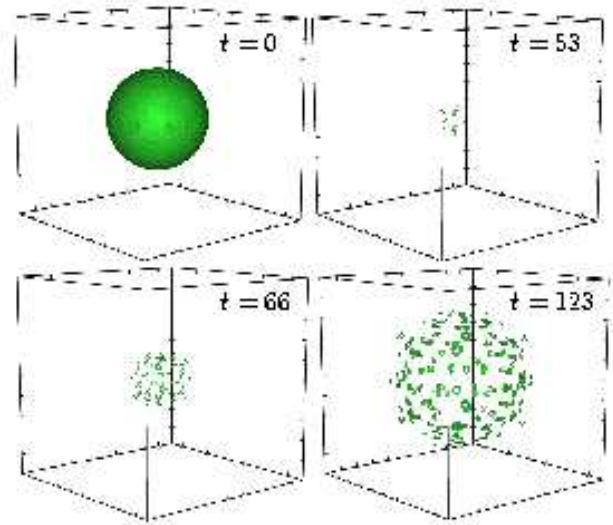, height = 3.1in}
\end{figure}

So far we considered the collapse of the stationary bubble, where the
vortex nucleation is connected to the instabilities developed in the
spherically symmetric flow. There are situations when the nucleation
is facilitated by an initial lack of the symmetry in the flow as in
the case of a moving bubble or a bubble in the nonuniform (trapped)
condensate. The surrounding 
helium exerts a net inward pressure across the surface, which is balanced 
by the  pressure  inside the bubble. In was shown in
\cite{br8} by asymptotic analysis of the GP equation coupled with the
equation of the motion for the wavefunction of an electron that a moving
bubble becomes oblate in the direction of its motion. This flattening is created by the 
difference in pressure between the poles and equator associated with the greater 
condensate velocity at the latter than at the former.  How oblate  the bubble becomes during its
motion will depend on the velocity and pressure inside the bubble. The
non-uniformity of the flow in the collapsing oblate bubble leads to
vortex ring nucleation  for bubble sizes much smaller than in the case
of a stationary spherically symmetric bubble. Figure 5 shows 
snapshots of the 
density plots of the cross-section of the collapsing bubble that prior
to $t=0$ was moving with a constant velocity $U=0.2$ and acquired an
oblate form given by $x^2 + \tfrac{1}{2}(y^2 + z^2) = 100$.  As the
result of  bubble's collapse four vortex rings of different radii
were created. Three of them are moving in the same direction as the
bubble before the collapse and one vortex ring is moving in the
opposite direction. In the model used, the oblateness of a
 bubble and
the velocity of its propagation, $U$, are independent parameters. A useful
quantity that can be determined from further numerical simulations is
the critical radius of the bubble as a function of the oblateness and
$U$. This will be further analysed
in our future research. 

In summary, we have suggested and studied a new mechanism of vortex 
nucleation as a result of the collapse of stationary and moving
bubbles in the context of the GP equation. 
This is related to the experiments in superfluid helium in which the
cavitated bubbles are generated by ultrasound in the megahertz frequency range. Our findings suggest that a sufficiently large or deformed stationary
bubble in a trapped condensate will produce vortex rings as a result
of its collapse. This will also be a subject of our future studies.
\begin{figure}
\caption{(colour online) 
The snapshots of the contour plots of the density cross-section of a
condensate obtained by numerically integrating the GP model
(\ref{gp}) for a moving bubble (see text). 
Black solid lines show zeros of real and imaginary parts of
$\psi$, therefore, their intersection shows the position of
topological zeros.  Both low and high density regions are shown in
darker shades to emphasise  intermediate density regions. Only a portion of an actual computational box is shown.
}
\centering
\bigskip
\bigskip
\epsfig{figure=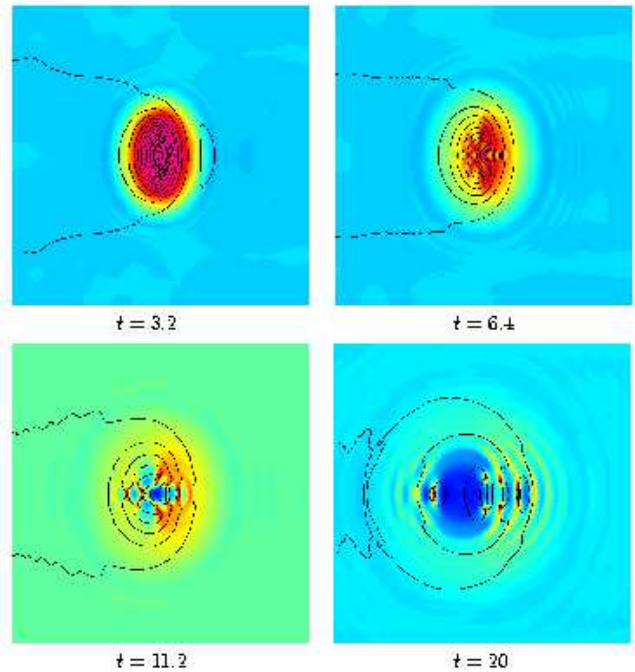, height = 3.6 in}
\end{figure}

NGB is supported by the NSF grant DMS-0104288; CFB is supported by
EPSRC grant GR/R53517/01.

\end{document}